\documentclass[aps,prl,twocolumn,preprintnumbers,noshowpacs,10pt]{revtex4-1}

\usepackage{amsmath}
\usepackage{amsfonts}
\usepackage{amssymb}
\usepackage{bbm}
\usepackage{hyperref}

\begin{document}
\title{Torus partition functions and spectra of gauged linear sigma models}
\preprint{DESY-14-027, LMU-ASC 11/14}
\author{Stefan \surname{Groot Nibbelink}}
\affiliation{Arnold-Sommerfeld-Center for Theoretical Physics, Fakult\"at f\"ur Physik, Ludwig-Maximilians-Universit\"at M\"unchen, Theresienstra\ss e 37, 80333 M\"unchen, Germany}
\author{Fabian \surname{Ruehle}}
\affiliation{Deutsches Elektronen-Synchrotron DESY, Notkestrasse 85, 22607 Hamburg, Germany}
\begin{abstract}
Worldsheet (0,2) gauged linear sigma models are often used to study supersymmetric heterotic string compactifications with non-trivial vector bundles. We make use of supersymmetric localization techniques to determine their one-loop partition functions. In particular we derive conditions which ensure that the full partition function is modular invariant and we propose a method to determine the massless and massive target space matter spectrum. 
\end{abstract}
\maketitle

\section{Introduction}

Ever since its introduction, string theory has inspired interesting progress, ranging from abstract mathematical insights to concrete phenomenological motivations for cosmology and particle physics of the standard model (SM) and beyond. String theory may be defined as a 2D conformal field theory (CFT) on the string worldsheet (WS) that maps {into} a target space, which is assumed to be 4D Minkowski space times a compact geometry (we focus on the heterotic string \cite{Gross:1985fr}).  As this leads to complicated non-linear sigma models, string phenomenology is often studied in its supergravity limit, compactified on special holonomy manifolds (e.g.\ Calabi-Yau (CY)) with vector bundles \cite{Candelas:1985en}, yielding supersymmetric grand unified or SM gauge groups and matter spectra.

As supergravity only constitutes the low energy effective field theory target space description of string theory, important WS quantum effects (e.g.\ Wilson line consistency conditions \cite{Blaszczyk:2009in2}) may be missed. Therefore, it is not obvious that every supergravity background has a lift to a full string theory. Due to their implicit definition (e.g.\ the metrics are unknown), descriptions of CYs with vector bundles mostly rely on topological methods. This makes computing the massless matter spectra a challenging task. Such complications are avoided by using orbifolds which admit exact CFT descriptions \cite{Dixon:1985jw,Ibanez:1987pj}. However, these geometries only account for special points in the moduli spaces of a minor set of all known CYs. 

Gauged linear sigma models (GLSMs) \cite{Witten:1993yc,Distler:1995mi} provide an approximate WS description for CY compactifications. Albeit not CFTs by themselves,  GLSMs are commonly assumed to flow to genuine CFTs in the infrared \cite{Silverstein:1994ih}. Given that these are again interacting WS theories, exact string results seem once more out of reach. However, using localization techniques \cite{Witten:1988ze} it is possible to determine their elliptic genera \cite{Grassi:2007va,Gadde:2013dda,Benini:2013nda,Benini:2013xpa,Gadde:2013lxa}. (Similarly, sphere partition functions \cite{Benini:2012ui,Doroud:2012xw} are related to exact moduli K\"ahler potentials \cite{Jockers:2012dk,Gomis:2012wy}.)

We go beyond these results and compute the full torus partition function by using only the globally Q-exact terms for localization and subsequently solve the resulting free theory. This allows us to derive one-loop modular invariance conditions necessary to ensure a string lift of the supergravity theory.
Moreover, we propose a method to determine the full target space spectrum up to the right-moving GSO projection. As such our results may be relevant for applications of string theory to physics beyond the SM. A detailed exposition of the derivation of our results with applications will be provided elsewhere.

\section{(0,2) gauged linear sigma models}

A GLSM is characterized in terms of matter superfields coupled to gauge multiplets \cite{Witten:1993yc,Distler:1995mi}. In the (0,2) superspace formalism these superfields are functions of the WS coordinates $\sigma, \bar\sigma$ and the Grassmann variables $\theta^+,\bar\theta^+$. The minimal characterization of these superfields is given by their gauge, global left-moving non-R- and right-moving R-symmetry charges. The normalization of the R-charge is chosen such that $\theta^+$ has R-charge 1. We consider the following (0,2) matter superfields 
\[
\begin{array}{|c||c|c||c|c|}
\hline 
\text{Super-} 
& \multicolumn{2}{|c||}{\mathcal{Z}^a,~~ \text{Chiral}} 
& \multicolumn{2}{|c|}{\Lambda^\alpha,~~  \text{Fermi} } 
\\ 
\text{field} & 
~~~z^a~~~ & ~~~\psi^a~~~ & 
~~~\lambda^\alpha~~~ & ~~~f^\alpha~~~
\\\hline\hline 
U(1)_{\phantom{\text{R}}} & 
 q_a & q_a & 
 Q_\alpha & Q_\alpha 
\\
U(1)_\text{R} & 
 r_a & r_a\text{-}1 & 
 R_\alpha & R_\alpha \text{-} 1 
\\
U(1)_\text{L\,} & 
 l_a & l_a  & 
 L_\alpha & L_\alpha 
\\ \hline 
\end{array}
\]
where $a=1,\dots,d$ and $\alpha=1,\dots,D$ label the chiral and Fermi superfields, respectively. 
In addition, we introduce the following bosonic and fermionic gauge multiplets 
\[
\begin{array}{|c||c|c|c||c|c|}
\hline 
\text{Super-} 
& \multicolumn{3}{|c||}{\mathcal{V},\mathcal{A},~ \text{Bos.\, gauge}~~~~} 
& \multicolumn{2}{|c|}{\Sigma,~ \text{Fermi\,gauge}} 
\\ 
\text{field} & 
~~~\mathfrak{a}~~~ & ~~~\chi~~~ & ~~~D~~~ & 
 ~~~~\varphi~~~~ & ~~~~\mathfrak{s}~~~~ 
\\\hline\hline 
U(1)_\text{R} & 
 0 & \text{-}1 & 0 & 
0 & \text{-}1 
\\ \hline 
U(1)_\text{L} & 
 0 & 0 & 0 & 
\text{-}1 & \text{-}1 
\\ \hline 
\end{array}
\]
where $\mathfrak{a}$ is a gauge field one-form. The fields $f^\alpha$ and $D$ are auxiliary. The fermionic gauge transformations read
\begin{equation}
\delta \Lambda^\alpha = \Xi\, N^\alpha(\mathcal{Z})~, 
\quad 
\delta \Sigma = \Xi~, 
\end{equation}
with the Fermi superfield $\Xi$ as gauge parameter. In the following, we allow for $D_\text{b}$ bosonic and $D_\text{f}$ fermionic gaugings labeled by $k$ and $\kappa$, respectively, and introduce the shorthand notation $q\cdot \mathfrak{a} = q_k \mathfrak{a}_k$.

Apart from the standard kinetic actions for the matter and gauge superfields the theory is specified by 
\begin{equation}
\mathcal{W} =  
G_\alpha(\mathcal{Z})\, \Lambda^\alpha~. 
\end{equation} 
This (0,2) superpotential has to be gauge and $U(1)_\text{L}$ invariant, have R-charge $1$ and be subject to $G_\alpha N^\alpha$=$0$. To encode target space K\"ahler deformations, Fayet-Iliopoulos (FI) parameters $\xi$ are included for the gauge multiplets. Absence of one-loop FI divergences requires
\begin{equation} \label{noFIren} 
\sum_a (q_k)_a = 
e_d^T q_k = 0~,
\qquad 
 e_d^T = (1,\ldots, 1)~. 
\end{equation}

\section{Zero modes} 

On the one-loop WS torus $T^2$ a gauge background $\mathfrak{a}$ is fully determined by its holonomies 
\begin{equation} 
\nu =  \tau\, a + a'~, 
\quad 
a = \oint_{\mathcal{C}_\tau} \mathfrak{a}~, 
\quad 
a' =  \oint_{\mathcal{C}_1} \mathfrak{a}~, 
\end{equation}
w.r.t.\ the cycles $\mathcal{C}_\tau$ and $\mathcal{C}_1$ in the fundamental directions of the WS torus with complex structure $\tau$. The holonomies take values on the same $T^2$, i.e.\ they are defined modulo the periodicities 
$\nu \equiv \nu + 1 \equiv \nu + \tau$. 
Similarly, the zero modes of the external gauge fields $\mathfrak{a}_\text{I}$, for $\text{I}=\text{L,R}$,  can be characterized by their holonomies $\nu_\text{I}= \tau\, a_\text{I} + a_\text{I}'$. 
The spin structures can be introduced by fixing the L- and R-symmetry holonomies to
\begin{equation} \label{SpinStructures} 
\nu_\text{R} = \tau\, \frac s2 + \frac {s'}2~, 
\qquad 
\nu_\text{L} = \tau\, \frac t2 + \frac {t'}2~, 
\end{equation} 
labeled by $s,t,s',t' =0,1$ for $\mathit{Spin}(32)/\mathbbm{Z}_2$. 
(The $E_8\times E_8$ theory requires two L-symmetries.)

The gauge holonomies $\nu$ are always zero modes. For specific values of the holonomies $\nu$ and $\nu_\text{L,R}$ further zero modes occur: 
\[
\begin{array}{|l||l|}
\hline 
\multicolumn{2}{|c|}{\text{Zero mode conditions}} 
\\\hline\hline 
\chi_0 & \nu_\text{R}=0
\\\hline 
\varphi_0 & \nu_\text{L}=0 
\\\hline 
\mathfrak{s}_0 & \nu_\text{L}=\nu_\text{R}=0 
\\\hline 
z_0^a & q_a \cdot \nu + l_a \nu_\text{L} + r_a \nu_\text{R} = 0
\\\hline 
\psi_0^a & q_a \cdot \nu + l_a \nu_\text{L} + (r_a-1) \nu_\text{R} = 0
\\\hline 
\lambda_0^\alpha & Q_\alpha \cdot \nu + L_\alpha \nu_\text{L} + R_\alpha \nu_\text{R} = 0
\\\hline 
\end{array} 
\]
For $\nu_\text{R}=0$ we encounter gaugino zero modes $\chi_0$ and the zero modes $\phi_0,\mathfrak{s}_0$ and $z_0^a,\psi_0^a$ always occur together.

\section{Supersymmetric localization}

The computation of path integrals of supersymmetric theories simplifies enormously due to localization~\cite{Witten:1988ze}. Assuming that the full action 
$S_\text{tot}(\Phi,\Psi) = (1/{g^2})\, S_\text{exact}(\Psi) + S_\text{rest}(\Phi,\Psi)$ 
can be split into an Q-exact part $S_\text{exact}(\Psi)$ of some superfields $\Psi$ w.r.t.\ a nilpotent operator Q$^2=0$ and a remainder Q-closed action $S_\text{rest}(\Phi,\Psi)$ of possibly additional superfields $\Phi$, the path integral
\begin{equation}
Z = \int {\cal D}\Psi {\cal D}\Phi\, e^{-S_\text{tot}}
\end{equation}
is, in fact, independent of the coupling constant $g$. Moreover, in the weak coupling limit $g \rightarrow 0$ one finds that 
\begin{enumerate}
\item[i)]
The classical zero mode background $\Psi_0$ localizes on
\begin{equation} 
S_\text{exact}(\Psi_0) = 0~. 
\end{equation} 
\item[ii)]  
The path integral over $\Psi$ reduces to a finite-dimensional zero mode integral: 
\begin{equation}
Z = 
 \int \text{d} \Psi_0~
 \text{Det}_{\delta \Psi}(\Psi_0) 
 \int {\cal D} \Phi\, 
e^{-S_\text{rest}(\Phi, \Psi_0)}~.  
\end{equation} 
\end{enumerate} 
The one-loop determinant factor $\text{Det}_{\delta \Psi}(\Psi_0)$ is obtained by computing the free partition function of canonically normalized quantum fluctuations $\delta \Psi$ after expanding 
\begin{equation}
\Psi = \Psi_0 + g\, \delta \Psi~, 
\quad 
\delta \Psi = \sum_{m,n} \Psi^m_n \, Y^m_n~, 
\end{equation}
in zero modes $\Psi_0$ and the torus mode functions $Y^m_n$.

\section{One-loop GLSM partition function} 

We want to apply supersymmetric localization to the GLSM path integral 
\begin{equation}
Z = \int {\cal D}(\mathcal{V},\mathcal{A}) {\cal D}\Sigma{\cal D} \mathcal{Z} {\cal D} \Lambda\, 
e^{-S}~,  
\end{equation} 
using the combination of supercharges  $\text{Q} = \text{Q}_+ + \overline{\text{Q}}_+$. However, localization arguments can only be applied when the supercharge Q defines an \textit{exact} symmetry of the theory. There might be various obstructions to this: 

Even when the theory is formulated in a superspace language, the torus boundary conditions for the bosons and fermions might be incompatible with supersymmetry. 
Hence, worldsheet Q-exactness necessitates restricting to the sector $\nu_\text{R}\equiv0$. 

Moreover, if the effective target space geometry $\mathcal{M}$ encoded in the bosonic zero modes $z_0$ of the chiral superfields is non-trivial, there is a global obstruction to Q-exactness~\cite{Hori:2001ax,Ashok:2013zka,Ashok:2013pya}. For this reason we only use the genuine Q-exact kinetic actions of the gauge, Fermi and Fermi gauge multiplets for localization. 
Even though the kinetic terms of the chiral superfields are not Q-exact globally, the chiral superfields become free fields when the localization of the Q-exact kinetic terms of the other superfields has been used. Consequently, the whole partition function can be computed exactly due to the interplay of localization and calculability of free field theories, which yields in the localization limit:
\begin{subequations} 
\begin{gather}
\label{D-term} 
D_0 \sim \sum_a q_{a} \big|z_0^a \big|^2 - \xi_\text{ren} = 0~,  
\\ 
\overline{f}_0^\alpha \sim  G_\alpha\big(z_0\big) = 0~,   
\qquad 
 \Big| \mathfrak{s}_0  \cdot N^\alpha\big(z_0  \big) \Big|^2  \sim 0~, 
\end{gather} 
\end{subequations} 
i.e.\ the well-known conditions on the target space geometry \cite{Witten:1993yc}. 
Even though the FI-action is Q-exact, there is nevertheless a dependence on the ratios of the FI-parameters $\xi$ and the normalization of the chiral superfield kinetic terms (which are not globally Q-exact) denoted by $\xi_\text{ren}$ in \eqref{D-term}.

In order to absorb the gaugino zero modes $\chi_0$, which occur in our calculations in the sector $\nu_\text{R}=0$, one expands the action $S_0$ to an appropriate order in $\chi_0$. As observed in \cite{Benini:2013nda}, the resulting expression can be written as $\overline\nu$-derivatives; the gauge holonomy integrals become contour integrals around the points 
$q_a \cdot \nu + l_a \nu_\text{L}  = 0$ in $T^2$ fundamental domains. Using a residue theorem, the holonomy integrals reduce to finite sums in this sector: The gauge holonomies $\nu$ get discretized.

\section{One-loop determinant factors}

In the supersymmetric sector $\nu_\text{R} \equiv 0$ all non-zero modes decouple in the limit $g\rightarrow 0$, so that their one-loop determinant factors become infinite products, like 
$\prod_{m,n} [m \tau + n + \omega]$ with $\omega =v\tau + v'$. Such infinite products can be $\zeta$-regularized, yielding expressions involving Dedekind $\eta$ and higher genus theta functions
\begin{equation}
Z_d\big[^v_{v'}\big] 
= \frac{\theta_1(\omega|\tau)}{\eta^d} 
= \frac{\theta\big[^{\frac 12 e_d -v}_{\frac 12 e_d -v'}\big]}{\eta^d}~. 
\end{equation} 
The resulting partition functions for chiral and Fermi superfields read 
\begin{subequations} 
\label{Dets} 
\begin{equation}\label{ChiralFermiDets} 
Z_\text{chiral} = 
\frac{\overline{ Z_d\big[^{v}_{v'}\big] }}
{\big|Z_d\big[^{v}_{v'}\big] \big|^2}~, 
\quad 
Z_\text{fermi} =Z_D\big[^{V}_{V'}\big]~, 
\end{equation}
with 
$v = q\cdot a + l \,a_\text{L}$ and $V=Q\cdot a+L\, a_\text{L}$ 
(and similarly for the primed versions). 
For the bosonic and fermionic gauge superfields we have 
\begin{equation} \label{GaugeDets}
Z_\text{bos.\,g.} = \Big( \tau_2 \eta^2(\tau) \Big)^{D_\text{b}}~,
\quad 
Z_\text{fer.\,g.} = \frac {\overline{Z_{1}^{D_\text{f}}\big[^{\text{-}a_\text{L}}_{\text{-}a_\text{L}'}\big]}}{\big|Z_{1}^{D_\text{f}}\big[^{\text{-}a_\text{L}}_{\text{-}a_\text{L}'}\big]\big|^{2}}~.
\end{equation} 
\end{subequations} 
In these expressions the anti-holomorphic dependence, which is due to WS supersymmetry,  drops out. Nevertheless, we prefer to keep them in this form in order to facilitate the spectrum determination later.

\section{Instanton contributions}

When the effective target space is compact the zero modes $z_0$ need not be constant but only harmonic, i.e.\ 
\begin{equation}
z_0(\sigma,\bar\sigma) = h_w(\sigma) + \widetilde{h}_w(\bar\sigma)~.
\end{equation}
The holomorphic functions $h_w(\sigma)$ define non-trivial mappings $T^2 \rightarrow \mathcal{M}$ characterized by sets of integers $w$. 
When the target space contains a torus, they encode the Kaluza-Klein and winding numbers. 
In the path integral we need to integrate over all possible field configurations, which leads to a lattice sum over the various instanton sectors labeled by $w$: 
\begin{equation}
Z_\text{inst}  = \text{Vol}(\mathcal{M})\, \sum_w e^{- S_{\text{ch\, kin}}(h_w,\widetilde h_w)}~, 
\end{equation}
where $S_{\text{ch\, kin}}$ is the kinetic energy of the zero modes of the bosons of the chiral superfields. The volume $\text{Vol}(\mathcal{M})$ of the target space manifold arises from the integral over the constant zero modes in $z_0$. In general, such lattice sums can be cast in the form 
\begin{equation}
Z_\text{inst} = \sum_{W \in \Gamma}\,  
q^{\frac 12\, W_\text{L}^2}\, \bar q^{\,\frac 12\, W_\text{R}^2} 
\end{equation}
via a partial Poisson resummation,
where $q = e^{2\pi i\, \tau}$and $W_\text{L,R}$ encode the mappings $h_w$. 
This partition function encodes the moduli dependence which arises because the chiral kinetic and superpotential actions are not globally Q-exact.

\section{Modular invariance}

Next we investigate the consequences of one-loop modular invariance of the full partition function 
\begin{equation}
Z = \int_{T^2} \text{d}^2\nu\, 
 Z_\text{dets}\, Z_\text{inst}~. 
\end{equation}
For simplicity we assume that the instanton partition function $Z_\text{inst}$ is modular invariant by itself, i.e.\ the lattice $\Gamma$ is even and self-dual. Simply taking $Z_\text{dets}$ as the product of the expressions in \eqref{Dets} does not produce a modular invariant result in general.  However, it is well-known \cite{Kawai:1986ah,Antoniadis:1987wp,Scrucca:2001ni} that by using so-called vacuum phases one can obtain expressions for partition functions,
\begin{equation}
\widehat Z_d\big[^v_{v'}\big] = e^{- \pi i v^T(v'-e_d)}\, Z_d\big[^v_{v'}\big]~,
\end{equation} 
that are modular invariant up to phases: 
\begin{subequations} 
\begin{eqnarray}
\widehat Z_d\big[^v_{v'}\big](\tau+1) 
&=&  e^{2\pi i \frac d{12}}\,  
\widehat Z_d\big[^v_{v'+v}\big](\tau)~, 
\\[1ex] 
\widehat Z_d\big[^v_{v'}\big](\tfrac {\text{-}1}\tau) 
&=& e^{-2\pi i \frac d4}\,\widehat Z_d\big[^{v'}_{\text{-}v}\big](\tau)
~. 
\end{eqnarray}
\end{subequations} 
Hence, by including such phase factors, we have a modular invariant partition function 
\begin{equation} \label{Zdets} 
\widehat Z_\text{dets} =
 \widehat Z_\text{chiral} \,
 \widehat Z_\text{fermi} \, 
 \widehat Z_\text{bos.\,g.}\, 
 \widehat Z_\text{fer.\,g.}~, 
\end{equation}
provided that e.g.\ $D-D_\text{f}= 12 + d - D_\text{b}$.
In order to reduce to the standard heterotic string in the absence of gaugings, we take $d=4+D_\text{b}$ and $D=16+D_\text{f}$.

This partition function is well-defined provided that the WS torus periodicities of the L- and gauge holonomies are respected. For $k, k' \in \mathbbm{Z}^d$ we have 
\begin{subequations} 
\begin{eqnarray}
\widehat Z_d\big[^{v+k}_{v'}\big] 
&=&  e^{\pi i k^T(e_d - v')}\,  
\widehat Z_d\big[^v_{v'}\big]~, 
\\[1ex] 
\widehat Z_d\big[^v_{v'+k'}\big] 
&=& e^{\pi i (v-e_d)^Tk'}\,\widehat Z_d\big[^v_{v'}\big]
~. 
\end{eqnarray}
\end{subequations} 
Since two different types of phases appear in these equations, requiring invariance under the periodicities of the holonomies $a_{\text{L}}$ and $a_k$, leads to both linear 
\begin{subequations} \label{ModInvL} 
\begin{eqnarray}
\tfrac 12 (e_D^T Q_k- e_d^T q_k) &\equiv&0~,\label{ModInvL1}
\\[2ex] 
\tfrac 12 (e_D^T L -e_d^T l + D_\text{f}) &\equiv& 0~,\label{ModInvL2}
\end{eqnarray}
\end{subequations} 
(where $a \equiv b$ means $a-b\in\mathbbm{Z}$) and quadratic conditions 
\begin{subequations} \label{ModInvQ} 
\begin{gather} \label{ModInvQ1} 
\tfrac {a_\text{L}}2 (L^2\!-\!  l^2 \!+\! D_\text{f}) \equiv 0~,
\\[2ex] 
 \label{ModInvQ2}
\tfrac {a_k}2 (Q_k^T Q_m \!-\! q_k^T q_m)\equiv \tfrac {a_m}2 (Q_k^T Q_m \!-\! q_k^T q_m)\equiv 0~,
\\[2ex]
\label{ModInvQ3}
\tfrac {a_{k}}2 (Q_k^TL \!-\! q_k^Tl) \equiv 
\tfrac {a_{\text{L}}}2 (Q_k^TL \!-\! q_k^Tl) \equiv 0~. 
\end{gather}
\end{subequations} 
The quadratic constraints \eqref{ModInvQ2}--\eqref{ModInvQ3} are the same as the GLSM $U(1)_k\times U(1)_m$ and $U(1)_k\times U(1)_\text{L}$ anomalies (the latter are strict equalities though). Since we generically only need the $U(1)_\text{L}$ to introduce the spin structures $t,t'$, \eqref{ModInvQ1} only has to be enforced for $a_\text{L}=1/2$. The linear constraint \eqref{ModInvL1} is automatically fulfilled for a target space with vanishing first Chern class and a bundle with trivial second Stiefel--Whitney class.

\section{Spectra}

The conventional massless spectrum computation of GLSMs makes use of $\overline{Q}$-cohomology (see e.g.\ \cite{Kachru:1993pg,Adams:2009tt,Bertolini:2014dia}). We outline an alternative way to determine the spectrum of GLSMs. Since the localization can only be applied in the supersymmetric sector, $\nu_\text{R}\equiv0$, we only have access to the Ramond sector, i.e.\ target space fermions. However, since the theory is supersymmetric in target space this is sufficient to determine the masses of all particles. This method follows the calculation of the (orbifold) CFT spectrum by making a $q$-expansion of the full partition function. Depending on whether theta functions appear in the numerator or denominator, one uses their sum or product representations, respectively. 

In detail, the left- and right-moving masses read
\begin{subequations}
\begin{eqnarray}
M_\text{R}^2 &=& \tfrac 12 W_\text{R}^2 + \tfrac 12 p_\text{R}^2 -\tfrac 12 + \delta c + N_\text{R}~, 
\\[2ex] 
M_\text{L}^2 &=& \tfrac 12 W_\text{L}^2 + \tfrac 12 P_\text{L}^2 -1 + \delta c + N_\text{L}~, 
\end{eqnarray}
\end{subequations} 
in terms of the so-called shifted momenta 
\begin{equation} 
p_\text{R} = 
\Big(  
{}^{p_d +v - \tfrac 12 e_d} 
_{p^{\;}_{D_\text{f}} - (a_\text{L} + \tfrac 12) e^{\;}_{D_\text{f}}}
\Big) 
~,~~
P_\text{L} = 
\Big( 
{}^{P_D + V - \tfrac 12 e^{\;}_D} 
_{P_{D_\text{b}} - \tfrac 12 e^{\;}_{D_\text{b}}} 
\Big)
\end{equation}
with $p_d\in \mathbbm{Z}^d$ and $v, V$ defined below \eqref{ChiralFermiDets}. 
We also introduced the number operators $N_{\text{L},\text{R}}$ and the vacuum shift  
\begin{equation}
\delta c = \tfrac 12 \tilde v^T ( e_d - \tilde v) 
+ \tfrac {D_\text{f}}2 \tilde v_\text{f} (1 - \tilde v_\text{f}) 
- \tfrac {D_\text{b}+D_\text{f}}8~, 
\end{equation} 
where 
$\tilde v \equiv  v$ 
and 
$\tilde v_\text{f} \equiv a_\text{L}$ 
such that all components satisfy $0 \leq \tilde v_a < 1$. 

The spectrum is subject to various projection conditions: 
The sums over the spin structure $t'$ lead to a left-moving GSO-projection. 
The sum over the discretized values of $a'$ induces a projection which is very similar to the orbifold projection. 
And as usual the level matching condition $M_\text{L}^2 =M_\text{R}^2$ is obtained from the $\tau_1$-integral.
However, because we cannot apply the localization techniques to the non-supersymmetric sector $\nu_\text{R}\neq0$ on the worldsheet, we are not able to read off the right-moving GSO-projection. Hence, the target space spectrum is still subject to one further projection which one might attempt to implement by hand on top of our results.

\section{Acknowledgments}
\begin{acknowledgments}
We thank V.\ Kumar, D.\ Israel, I.\ Melnikov, F.\ Quevedo, R.\ Valandro and P.K.S.\ Vaudrevange for discussions and correspondence. 
The work of SGN was supported by the LMU Excellent Programme.
The work of FR was supported by the German Science Foundation (DFG) within the Collaborative Research Center (SFB) 676 ``Particles, Strings and the Early Universe''.
\end{acknowledgments}

%
%

\bibliographystyle{apsrev4-1}

\end{document}